\documentclass[aps,prl,twocolumn,superscriptaddress,twoside,floatfix,amsmath,showpacs,nofootinbib]{revtex4}

\usepackage{graphicx}
\usepackage{amsmath}
\usepackage{textcomp}
\usepackage{dcolumn}
\usepackage{bm}

\usepackage{amssymb}
\usepackage[latin1]{inputenc}
\usepackage{amsfonts}
\usepackage{lineno}
\usepackage{url}
\usepackage{array}
\usepackage{verbatim}
\usepackage{subfigure}
\usepackage{listings}
\usepackage{color}
\usepackage{lscape}
\usepackage{setspace}
\usepackage{footmisc}

\usepackage [english]{babel}
\usepackage [autostyle, english = american]{csquotes}
\MakeOuterQuote{"}

\begin{document}


\title{New source for ultracold neutrons at the Institut Laue-Langevin}

\author{F.M.~Piegsa}
\altaffiliation{Electronic addresses: florian.piegsa@phys.ethz.ch (FMP); zimmer@ill.fr (OZ).}
\affiliation{Institut Laue-Langevin, BP 156, F-38042 Grenoble, France}
\affiliation{ETH Z\"urich, Institute for Particle Physics, CH-8093 Z\"urich, Switzerland}

\author{M.~Fertl}
\affiliation{Institut Laue-Langevin, BP 156, F-38042 Grenoble, France}
\affiliation{Paul Scherrer Institute, CH-5232 Villigen PSI, Switzerland}

\author{S.N.~Ivanov}
\affiliation{Institut Laue-Langevin, BP 156, F-38042 Grenoble, France}

\author{M.~Kreuz}
\affiliation{Institut Laue-Langevin, BP 156, F-38042 Grenoble, France}

\author{K.K.H.~Leung}
\affiliation{Institut Laue-Langevin, BP 156, F-38042 Grenoble, France}

\author{P.~Schmidt-Wellenburg}
\affiliation{Institut Laue-Langevin, BP 156, F-38042 Grenoble, France}
\affiliation{Paul Scherrer Institute, CH-5232 Villigen PSI, Switzerland}

\author{T.~Soldner}
\affiliation{Institut Laue-Langevin, BP 156, F-38042 Grenoble, France}

\author{O.~Zimmer}
\altaffiliation{Electronic addresses: florian.piegsa@phys.ethz.ch (FMP); zimmer@ill.fr (OZ).}
\affiliation{Institut Laue-Langevin, BP 156, F-38042 Grenoble, France}





\date{\today}

\begin{abstract}
A new intense superthermal source for ultracold neutrons (UCN) has been installed at a dedicated beam line at the Institut Laue-Langevin. 
Incident neutrons with a wavelength of 0.89~nm are converted to UCN in a five liter volume filled with superfluid $^4$He at a temperature of about 0.7~K. The UCN can be extracted to room temperature experiments. We present the cryogenic setup of the source, a characterization of the cold neutron beam, and UCN production measurements, where a UCN density in the production volume of at least 55 per cm$^3$ was determined.



\end{abstract}












\maketitle

\section{Introduction}

\label{sec:introduction}

The neutron offers access to many different physical phenomena. 
Its static and decay properties as well as its interactions are 
studied at research reactors and spallation sources. High
precision experiments have impact on particle physics and cosmology and also serve for
sensitive searches for new physics, complementary to work done at
high-energy accelerators \cite{Dubbers/2011,Musolf/2008,Abele/2008}. For
several important investigations ultracold neutrons (UCN) have proven to be
particularly useful. The defining property of UCN is total reflection
under any angle of incidence by virtue of the neutron optical potential.
As a result one can confine them in "neutron bottles" made of suitable materials with
small cross sections for neutron absorption and trap depths of up to
about 300~neV. Good neutron bottles provide storage time constants of many
hundreds of seconds \cite{Golub/1991}. \\
Current fields of research with UCN include the search for a non-vanishing neutron electric dipole moment \cite{Baker/2006,Altarev/1996,Grinten/2009,Masuda/2012a,EDM@SNS/2004,Altarev/2012,Baker/2011,Serebrov/2009b,Lamoreaux/2009},
searches for "mirror matter" as a viable candidate for dark matter \cite{Ban/2007,Serebrov/2008}, 
sensitive tests of Lorentz invariance \cite{Altarev/2009,Altarev/2010},
searches for new fundamental forces mediated by axion-like particles \cite{Baessler/2007,Serebrov/2010,Zimmer/2010b},
measurements of the neutron lifetime \cite{Serebrov/2008b,Pichlmaier/2010,Arzumanov/2012,Huffman/2000,Paul/2009} and
angular correlation measurements in neutron $\beta$-decay \cite{Mendenhall/2013}.
UCN are also useful for investigations of quantum mechanical phenomena, e.g.\ quantum states of the neutron in the Earth's gravitational field \cite{Nesvizhevsky/2002,Jenke/2011,Jenke/2012}. The new source described here will provide UCN for the GRANIT experiment, a gravitational quantum level experiment, which is currently being set up at the Institut Laue-Langevin (ILL) in Grenoble, France \cite{Kreuz/2009,Wellenburg/2009a}. \\
The development of new sources has become strongly motivated by counting statistical limitations 
apparent in many of the aforementioned studies, which mostly were performed at the UCN source 
PF2 at ILL, consisting of a neutron turbine coupled to a liquid deuterium moderator \cite{Steyerl/1986}. 
In order to overcome these limitations
several laboratories have embarked in developing UCN sources of a "next generation" 
\cite{Trinks/2000,Masuda/2012,Saunders/2013,Korobkina/2007,Frei/2007,Serebrov/2009,Anghel/2009,Karch/2013,Lauer/2013}, 
all based on the "superthermal" UCN production scheme proposed by Golub and Pendlebury in 1975 \cite{Golub/1975}. 
In this concept neutrons incident on a cold converter made of solid deuterium or superfluid $^{4}$He 
loose almost their entire energy in single scattering events, producing
elementary excitations (phonons) in the converter medium which are cooled
away by a refrigerator. At low temperatures the probability for up-scattering
of UCN back to higher energies is suppressed by the Boltzmann
factor. In solid deuterium the survival time for free neutrons is only a
fraction of a second. For competitive UCN densities a sD$_{2}$ 
converter therefore needs to be implemented "in-pile" close to the
core of a reactor or near a spallation target. On the other hand, $^{4}$He
has no cross section for neutron absorption so that a viable concept for
production of competitive UCN densities places the converter at the end of a
neutron guide \cite{Golub/1977}. No extraordinary cooling power is required
there so that the source can be designed as a relocatable apparatus easy to
access for maintenance and trouble-shooting. UCN storage time constants in $^{4}$He 
are ultimately limited only by the neutron beta decay lifetime of
about 880~s \cite{PDG/2012,Wietfeldt/2011}. Therefore, albeit a neutron guide delivers a much lower flux
than available in-pile, UCN may be accumulated to high density in a
converter with reflective walls. The kinematics defined by the dispersion
relations of helium and the free neutron enables down-scatter of cold
neutrons with an energy around $1.0$ meV (corresponding to a neutron
wavelength of 0.89~nm) to ultracold energies via emission
of a single phonon. To a lower extent, also multi-phonon processes
contribute to the integral UCN production rate for a wide range of incident
neutron energies \cite{Korobkina/2002,Wellenburg/2009,Baker/2003}.\\
While an attempt to extract accumulated UCN horizontally from a
superfluid helium converter was hampered by large losses \cite%
{Kilvington/1987}, vertical UCN extraction through a cold mechanical valve
for UCN does not require any windows in the UCN guide connecting the
converter with an experiment at room temperature. Already some years ago the
concept was shown to solve the problem, using a prototype apparatus
installed at a neutron beam at the Munich research reactor FRM II \cite%
{Zimmer/2007,Zimmer/2010}.\footnote{Despite a difference of potential energy of about 102~neV per meter rise
of a neutron in the gravity field, a short vertical section of well polished
UCN guide is no obstacle for extraction if it does not exceed about 20~cm.
This is due to UCN getting boosted by the neutron optical potential of the
superfluid $^{4}$He when they leave through the surface of the bath. }
In continuation of that development with the goal to make the technique
available for the GRANIT experiment,
the apparatus was upgraded and implemented at the new beam line H172a at
ILL \cite{Wellenburg/2009a,Andersen/2010}. Results of a
commissioning run at the end of 2010 were already published \cite{Zimmer/2011}. 
Here, we give a detailed description of the source, the
measured characteristics of the beam, and present more results on UCN
production obtained with this improved device.

\section{The ultracold neutron source}
\label{sec:ucnsource}

\begin{figure*}[tbp]
\centering
\includegraphics[width=0.70\textwidth]{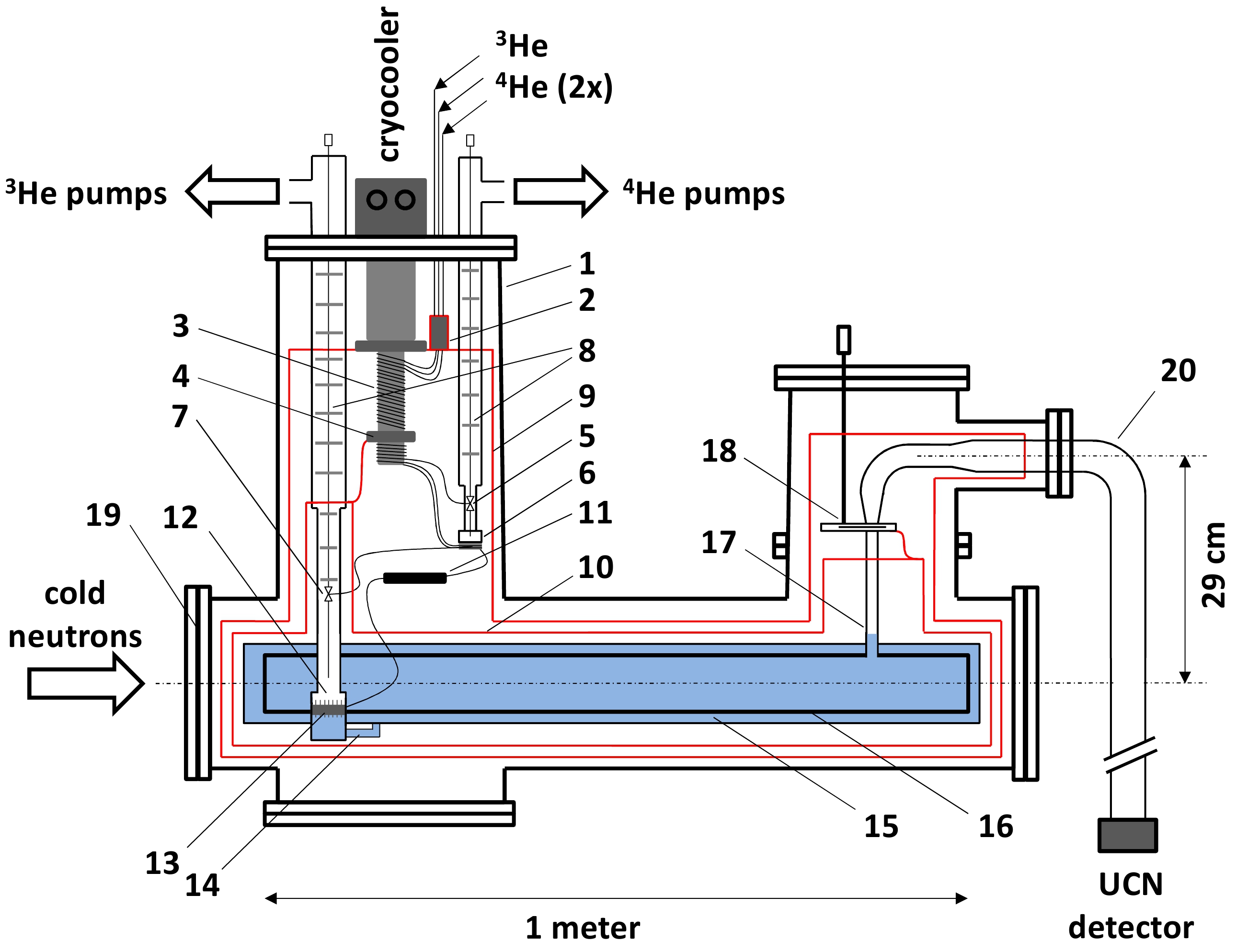}  
\caption{(Color online) Schematic drawing of the cryostat of the UCN source: (1) isolation
vacuum vessel, (2) coldtraps and heat-exchangers for incoming helium gas
thermally anchored to the first stage of the cryocooler at about 45~K, (3)
regenerator heat-exchangers, (4) second stage of the cryocooler with
heat-exchangers, (5) cold needle valve of the $^4$He circuit, (6) 1-Kelvin $^4$He 
evaporation pot, (7) cold needle valve of the $^3$He circuit, (8) thermal baffles,
(9) outer heat shield attached to the first stage of the cryocooler, (10)
inner heat shield attached to the second stage of the cryocooler, (11)
superleak filter, (12) $^3$He evaporation pot, (13) copper heat-exchanger, (14) "U"-shaped
connection pipe, (15) aluminum converter vessel filled with superfluid $^4$%
He, (16) UCN storage/production volume, (17)
vertical stainless steel UCN guide, (18) UCN shutter valve thermally 
anchored to the inner heat shield, (19) thin aluminum cold neutron beam window, and (20) stainless steel
UCN guide at room temperature.}
\label{fig:cryo}
\end{figure*}
\begin{figure}[tbp]
\centering
\includegraphics[width=0.45\textwidth]{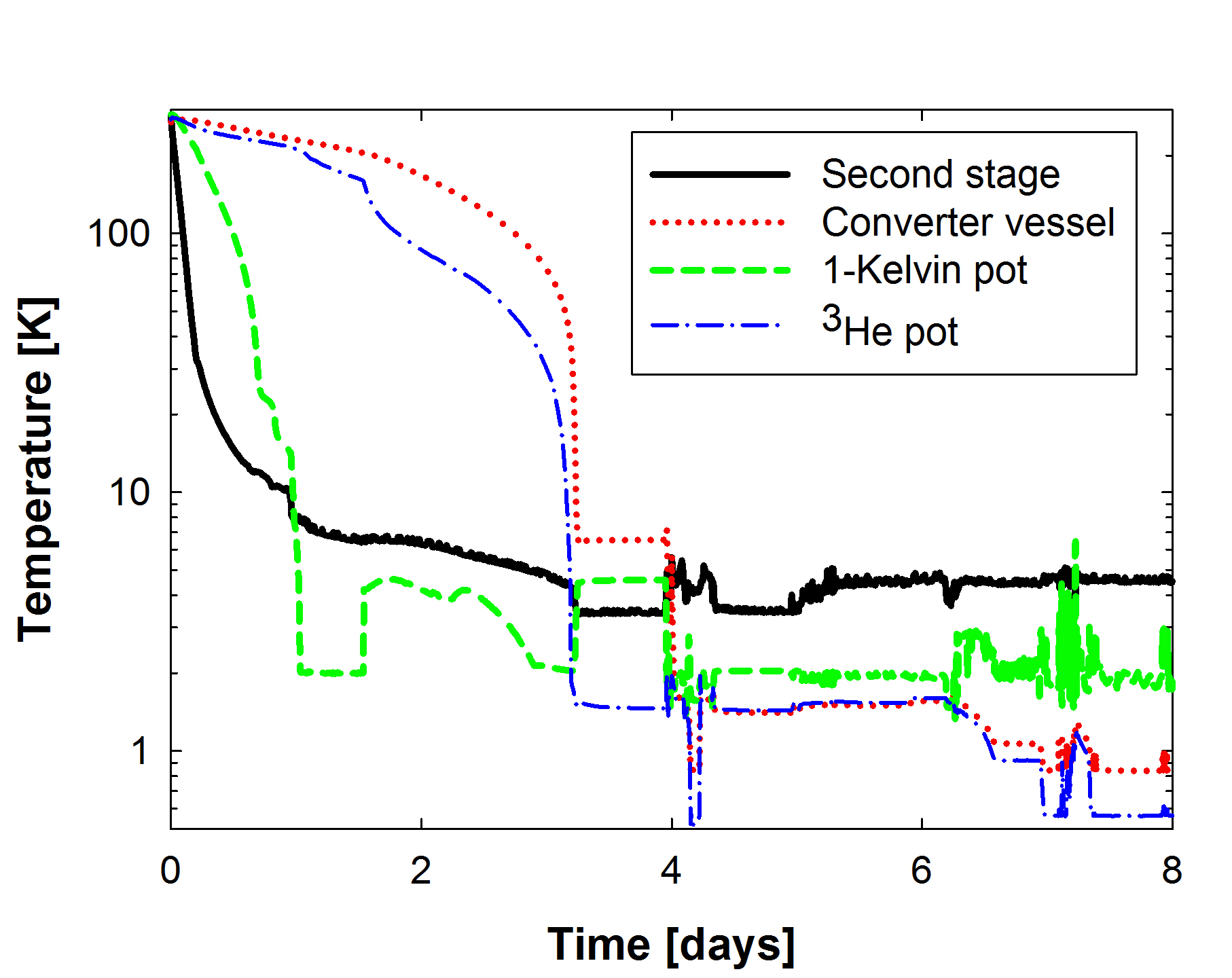}  
\caption{(Color online) Time plot of a cooldown of the UCN source cryostat starting from room temperature. After
about seven days the filled converter reaches a temperature of 0.85~K
(measured outside of the converter at the UCN extraction side, see text),
while the temperature in the $^3$He pot is
0.55~K.  }
\label{fig:cooldown}
\end{figure}

The key component of the presented superthermal UCN source is the cryostat
which cools a large aluminum converter vessel filled with approximately 7
liters of superfluid $^4$He to below 1~K. 
The converter liquid is cooled by coupling it thermally to a pumped $^3$He bath.
In Fig.\ \ref{fig:cryo} a schematic drawing of the complete cryogenic system
is depicted. All low temperature parts are enclosed in an isolation vacuum
vessel ($p_{\text{iso}} < 10^{-6}$~mbar) and two thermal heat radiation
shields produced from thin copper and aluminum sheets. Two separate closed 
cooling circuits using $^3$He and $^4$He are fed into the cryostat via two gas inlets. 
A third inlet serves to fill the converter vessel with $^4$He from a gas cylinder. 
All helium gas is filtered by passing it through charcoal cold-traps immersed in liquid nitrogen 
(located external to the cryostat and not shown in Fig.\ \ref{fig:cryo}) 
to prevent blocking of capillaries inside the cryostat by gas impurities.
A commercial Gifford-McMahon cryocooler (Sumitomo RDK-415E, 1.5~W at 4.2~K)
is used to precool and finally liquefy the gas via heat-exchangers
attached to the first and second stage of the
cryocooler, and to the regenerator between the two stages \cite{Wellenburg/2006}. 
For all temperature measurements calibrated Lakeshore Cernox
thermometers with a typical sensor accuracy of $\pm 5$~mK were used. 
\begin{figure}[tbp]
\centering
\includegraphics[width=0.45\textwidth]{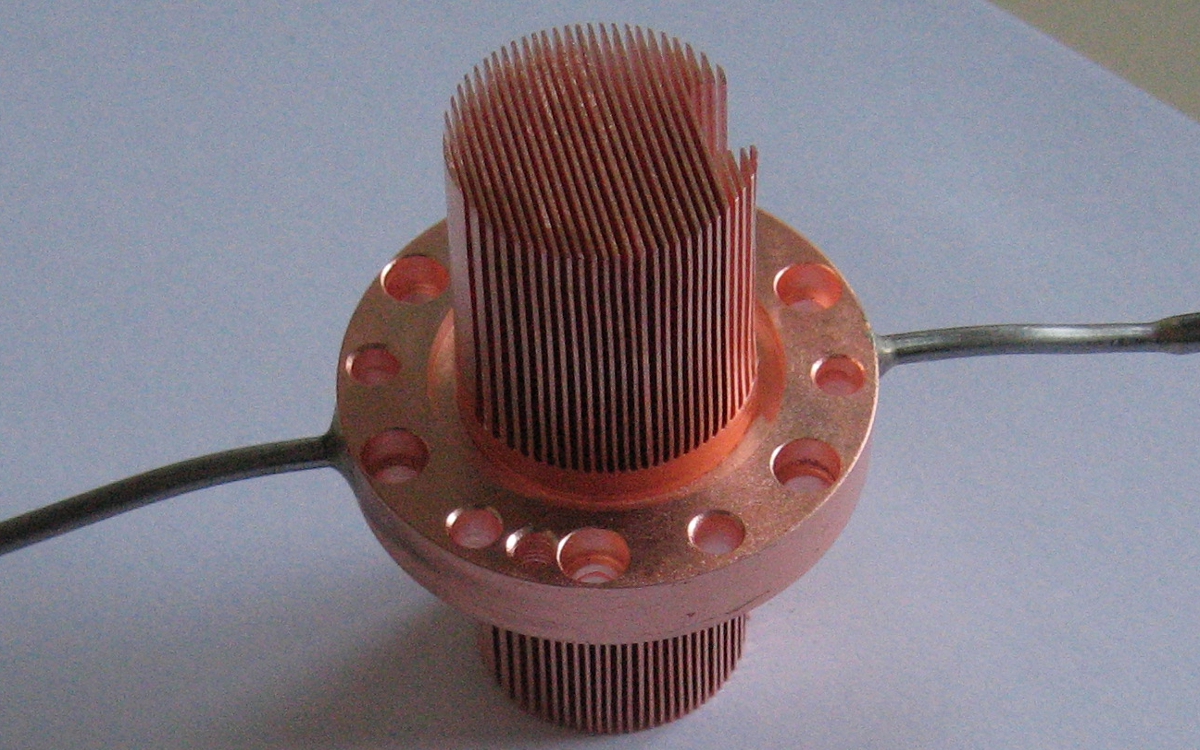}  
\caption{(Color online) Copper heat-exchanger between the $^3$He pot and the superfluid $^4$%
He bath in the converter vessel. The eroded fins (28~mm long and 0.5~mm thick)
result in a surface area of about 300~cm$^2$ on each side.}
\label{fig:HE_copper}
\end{figure}\newline
The $^4$He circuit is operated at an absolute inlet pressure of about $5-6$~bar 
employing a small membrane compressor (KNF - N145 AN 9E). Hence, the $^4$%
He gas can be liquefied on the second stage of the cryocooler 
even at elevated temperatures in the range $4.5-5.0$~K caused by the thermal load. The liquid $^4$He is then continuously transfered
into the so-called 1-Kelvin pot, which has a volume of about 40~cm$^3$. The flow
of liquid $^4$He is regulated using a cold needle valve to approximately
$30-40$~g/h. Two 40~m$^3$/h dry pumps (Alcatel ACP 40) are used in parallel to
pump on the 1-Kelvin pot achieving a temperature of $1.5-2.0$~K. This
temperature strongly depends on the flow rate of $^3$He, as the precooled $^3
$He gas coming from the second stage of the cryocooler is liquefied in a
heat-exchanger capillary situated inside the 1-Kelvin pot. The $^3$He circuit is operated
at a much lower absolute inlet pressure of about 0.8~bar. After being liquefied, the 
$^3$He is continuously transfered into the $^3$He pot. The flow of liquid $^3$He
is regulated by a cold needle valve to about $10-15$~g/h. By pumping with two
consecutive roots pumps (Pfeiffer, 2000~m$^3$/h and 500~m$^3$/h) backed by a
dry pre-pump (Alcatel ACP 40) a base temperature of 0.55~K is reached in
the $^3$He pot. The total amount of circulating $^3$He is about 1.5~g. \newline
The $^4$He gas used to fill the converter vessel is also precooled and liquefied on the
stages of the cryocooler and then further cooled below the superfluid transition temperature 
and filtered by passing a "superleak" thermally anchored to the
1-Kelvin pot \cite{Zimmer/2010}. This filtering is important to remove traces of the 
strongly neutron absorbing isotope $^3$He from the liquid.
Finally, the heat transfer between the $^3$He pot and the converter vessel is achieved by
a heat-exchanger produced from copper of high purity (a quality normally
used for single crystal neutron monochromator production), shown in Fig.\ \ref{fig:HE_copper}. 
The imperfect heat transfer of the heat-exchanger, due to the Kapitza resistance, 
results in a temperature gradient of about 0.15~K, which causes a slightly 
higher temperature of the superfluid $^4$He in the
converter of about 0.7~K (measured with a resistor thermometer attached on the
outside of the thin walled "U"-shaped copper pipe, item 14 in Fig.\ \ref{fig:cryo}). A thermometer 
of the same type mounted at the opposite end on the outside of the converter
vessel showed a higher temperature of 0.85~K, 
which might be due to a worse thermal coupling of the thermometer to the liquid helium 
(the thermometer was attached at a position above the helium bath, where the aluminum 
of the vessel has a thickness of approximately 1~cm and is not necessarily in direct contact with the superfluid).
\begin{figure}[tbp]
\centering
\subfigure[]{\includegraphics[width=0.45\textwidth]{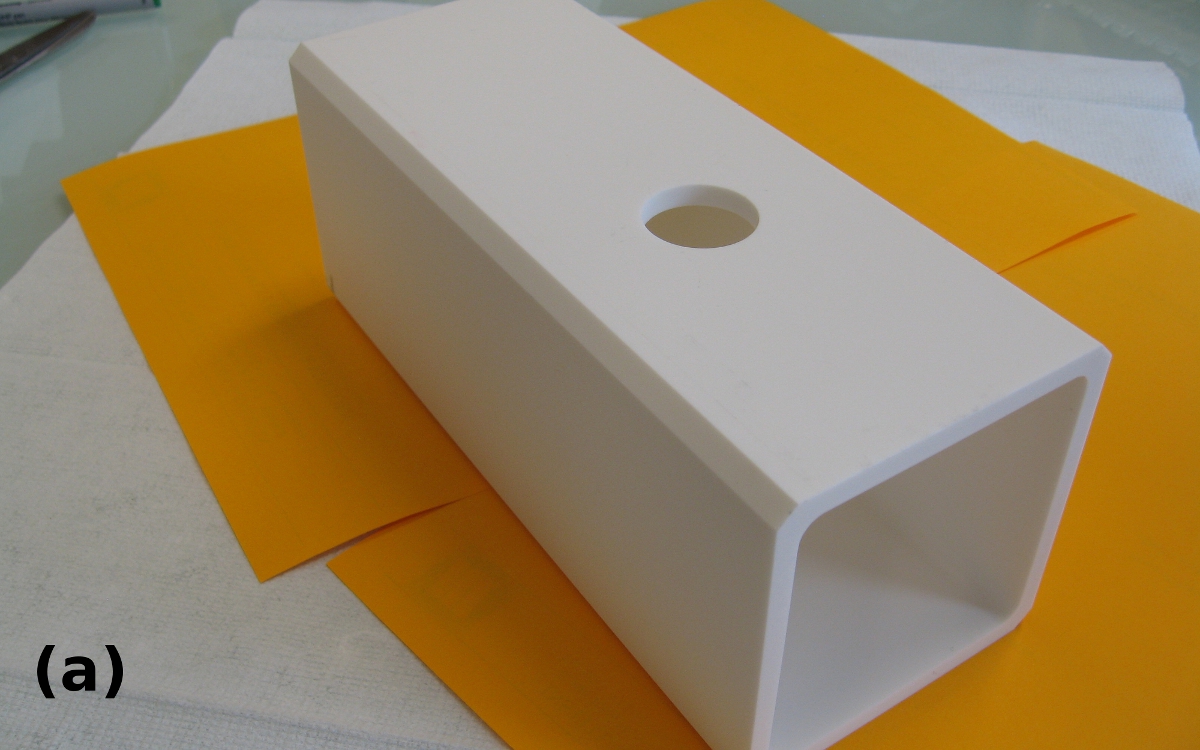}  } 
\subfigure[]{\includegraphics[width=0.45\textwidth]{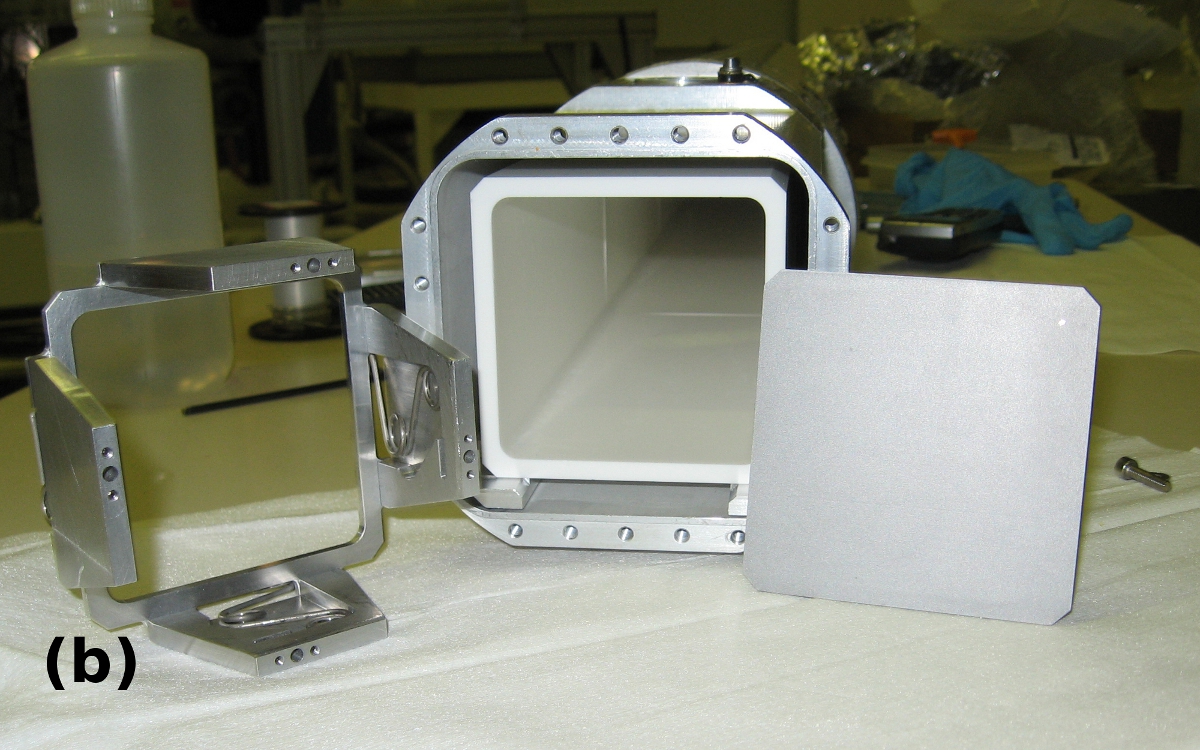}  } 
\subfigure[]{\includegraphics[width=0.45\textwidth]{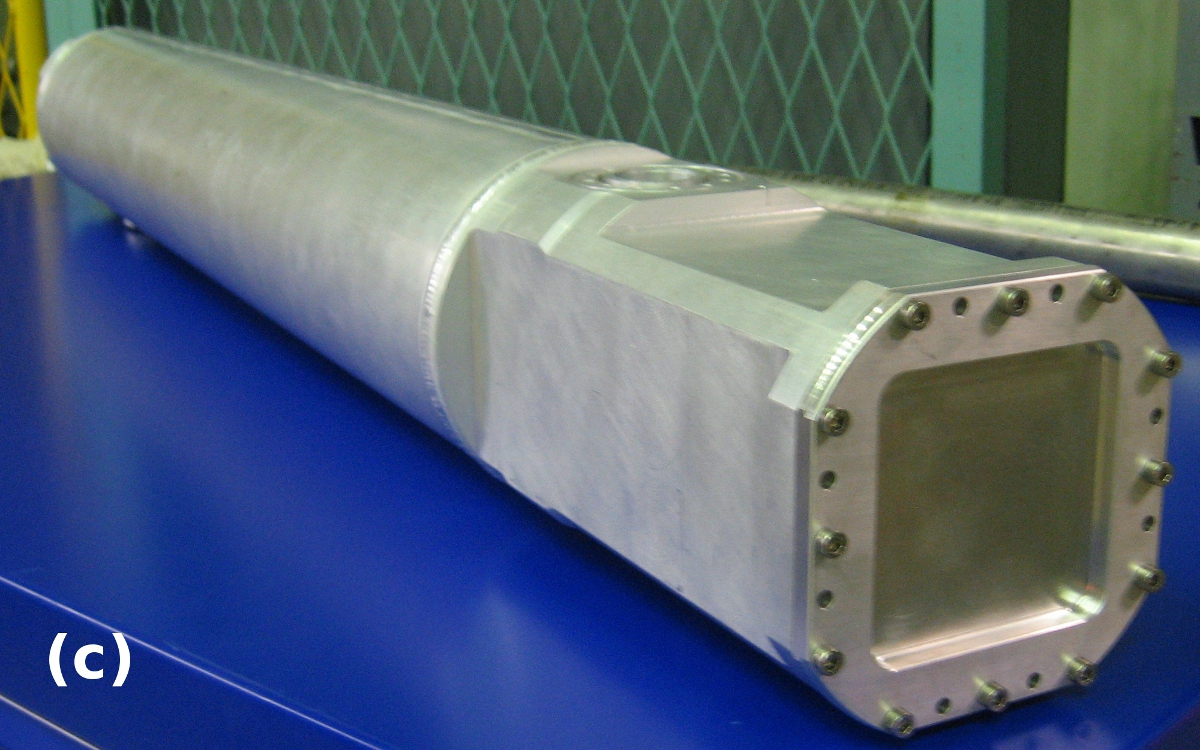}  }  
\caption{(Color online) (a) One of the five 20~cm long BeO ceramic pieces forming the 5
liter UCN storage volume. Its inner cross section is $70\times70$~mm$^2$.
The hole in the piece shown serves for extraction of the UCN via a stainless 
steel guide (see text). (b) Front view of the opened aluminum converter
vessel with the installed white BeO ceramic pieces. The storage volume is
closed off with two 1~mm thick beryllium sheets (right) serving as entrance and exit
windows for the cold neutron beam. The spring mechanism used to press 
the ceramic parts together is presented on the left. (c) Closed converter vessel with an
indium-sealed thin aluminum entrance window (approximate converter vessel
length: 1~m).}
\label{fig:volume_total}
\end{figure}  \newline
In order to store the produced UCN a 5 liter UCN storage/production volume 
with the inner dimensions $70 \times 70 \times 1000$~mm$^3$ is placed inside the converter vessel.
This volume is composed of five pieces of BeO ceramics and two 1~mm thick beryllium end-caps (compare Fig.\ \ref%
{fig:volume_total}). Both materials have large neutron optical potentials
(BeO: 261~neV and Be: 252~neV \cite{Golub/1991}) and low neutron absorption cross sections.
To avoid UCN losses through potential gaps due to thermal contraction, the set of BeO
ceramics is pressed together by a spring mechanism. \\
For UCN extraction, a guide system is connected to a hole with diameter 
24~mm in one of the BeO pieces. The entire UCN extraction guide system 
consisting of various tubing elements is made 
from polished stainless steel with a calculated neutron optical potential of
184~neV. The first, vertical section of the UCN extraction guide with 
inner diameter 23~mm is made to fit tightly into the hole in the BeO 
ceramics. Thirty millimeters below the end of this 173~mm long section a UCN 
shutter valve is located. The valve is made from a 0.5~mm thick sheet of stainless steel
with a hole (diameter 23~mm) which can be rotated to positions 
between closing and fully opening the guide, via a rotary feedthrough to room
temperature. A short conical section with half opening angle of 75° 
makes a transition to an 8~mm long straight section with inner diameter 50~mm, followed by 
a 90° bend with radius 70~mm connected with a horizontal, straight 
section of 342~mm length. A 65~mm long conical section increases the 
inner guide diameter from 50~mm to 66~mm. After an additional horizontal, 90~mm 
long straight section followed by a bend downwards with radius 80~mm
a vertical, straight section with inner diameter 66~mm and 
length 1.65~m is connected. At the end of this section, 
the entrance foil to a UCN detector is situated. 
This foil with thickness 100~$\mu$m is made from pure aluminum, 
followed by a 5 cm thick detection volume filled with 14~mbar of the 
strongly neutron absorbing gas $^3$He, 1.1~bar of Ar and 25~mbar of CO$_2$ as 
quench gas. The neutron detector is expected to have an efficiency close 
to 100\% for UCN which enter the gas detector volume.\\
The cooldown of the apparatus including filling of the converter 
takes approximately one week and is presented
in a time plot of 4 thermometers in Fig.\ \ref{fig:cooldown}. About 3
days after the cryocooler was started the converter vessel as well as the
1-Kelvin and $^3$He pots have thermalized below 10~K. During the whole time
helium is circulated in the two closed gas circuits to establish a good
thermal contact and fast cooling. Afterwards, the $^4$He is filled in the
converter vessel via the superleak (approximately 1~kg, 
visible in the plot as period when the temperatures of
the $^3$He pot and the converter vessel are equal). This takes approximately
another 2-3 days, before it is cooled from about 2~K to below 1~K within a
few hours with the cooling power of the $^3$He pot. 

\section{Characterization of the beam line H172a}
\label{sec:beamline}

The general outline of the new beam line has already been described in Ref.\ \cite{Wellenburg/2009a}. 
In the following section, we present its performance in detail.

\subsection{Neutron guide system and monochromator}

\begin{figure}[tbp]
\centering
\includegraphics[width=0.45\textwidth]{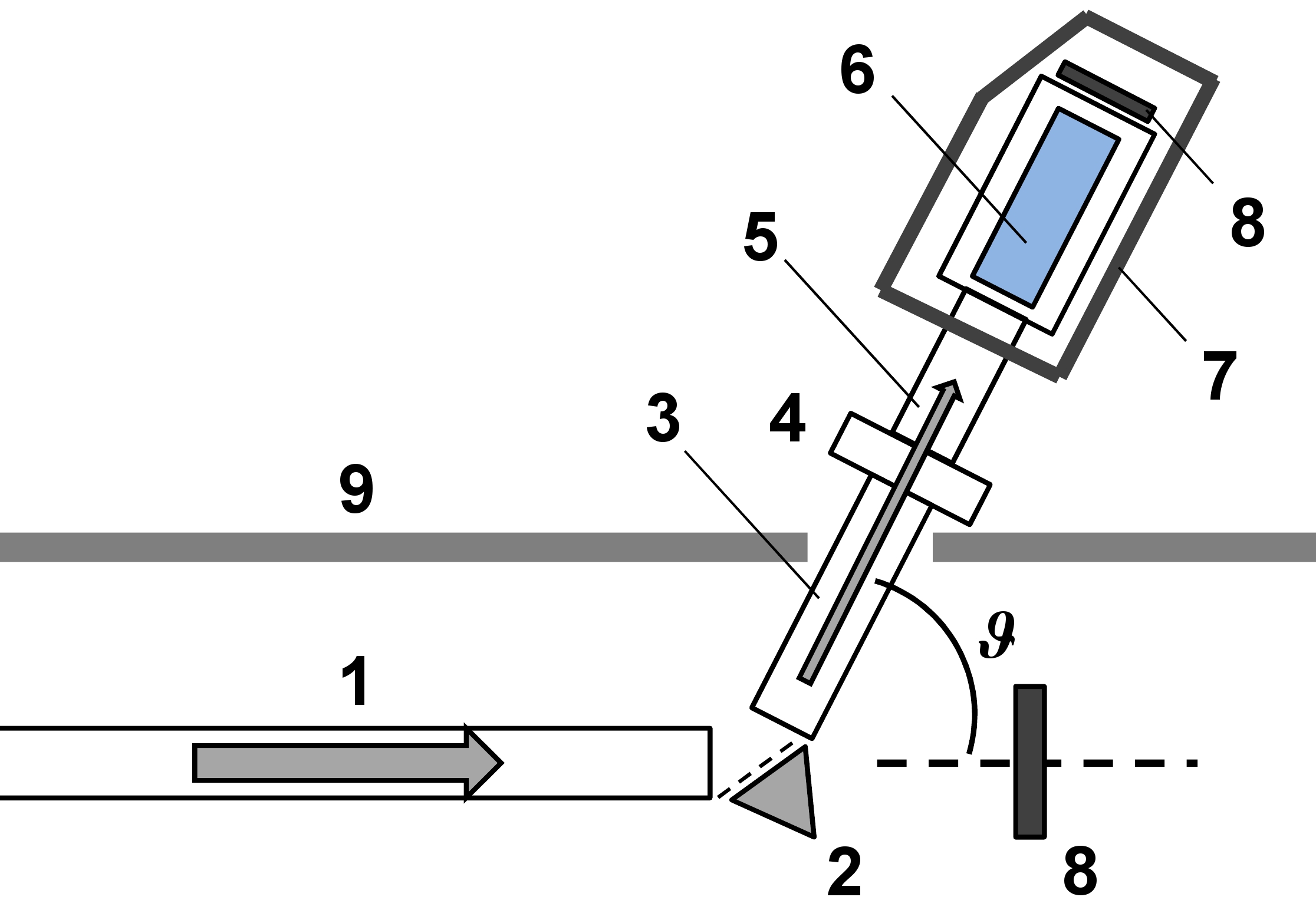}  
\caption{(Color online) Schematic top-view of the beam line: (1) neutron guide H172 coming from the vertical cold source of the ILL reactor, 
(2) monochromator crystals (dashed lines) attached to triangular shaped holder, 
(3) long neutron guide section of H172a, converging from $80\times80$~mm$^2$ to $70\times70$~mm$^2$ and 
tilted by $\protect\vartheta=61.2$° with respect to the incoming neutron beam, (4) secondary beam shutter, 
(5) short neutron guide section of H172a, (6) UCN source with helium filled converter vessel, 
(7) lead shielding with a thickness of 7~cm, (8) beam stops, and (9) concrete wall for biological shielding.}
\label{fig:beam}
\end{figure}

The superthermal UCN conversion mechanism in superfluid $^4$He is most
efficient within a narrow neutron wavelength band at 0.89 nm \cite{Baker/2003}. Placing
the UCN source in a monochromatic neutron beam reduces the created 
$\gamma$-radiation and activation with respect to a converter situated in a white
beam. In this case, multi-phonon processes will not contribute to UCN production
\cite{Wellenburg/2009,Korobkina/2002}. A top-view drawing of the
beam line is depicted in Fig.\ \ref{fig:beam}. 
Neutrons from the white beam of neutron
guide H172 coming from the vertical cold source of the ILL high-flux
reactor are deflected by $\vartheta=61.2$° by means of a monochromator. They are fed into a
secondary 4.5~m long neutron guide (H172a, items 3 and 5 in Fig.\ \ref{fig:beam}) 
equipped with a $m=2$ supermirror coating.\footnote{"$m=2$" stands for a coating which 
provides a critical angle of total reflection two
times larger than for nickel with natural isotopic composition.} The
secondary neutron guide is tapered to reduce the cross section from $80
\times 80$~mm$^2$ to $70 \times 70$~mm$^2$ matching the cross section
of the BeO ceramic pieces of the UCN storage volume. \\
The cold neutron beam incident on the UCN converter vessel can be switched on/off by means of a beam shutter, 
consisting of a 187~mm long block of lead, covered with 13~mm of B$_4$C, 
with a switching time of about 1~s (item 4 in Fig.\ \ref{fig:beam}). 
When opening the shutter, a piece of neutron guide integrated into the block is moved
into the beam which avoids losses due to beam divergence. \\ 
The monochromator is composed of 18 potassium intercalated graphite crystals of
dimensions $45\times20\times3$~mm$^3$. Such crystals can be manufactured, depending on production conditions, to
possess a structure with layers of potassium atoms situated between each
layer of carbon atoms ("stage-1" type crystal, with a lattice spacing $d_{\text{1}}=0.535$~nm) or
between each second carbon layer ("stage-2" type crystal, $d_{\text{2}}=0.875$~nm). 
The beam line described here employs stage-2 crystals 
to scatter a 0.89~nm neutron beam by a first order Bragg
reflection under the angle $\vartheta$. The crystals have a
peak reflectivity of $60-70$\% and a mosaicity of about 2°, which matches the
divergence of the incident beam \cite{Courtois/2011}. 
They are mounted with a gap of approximately 2~mm between two crystals, 
resulting in an estimated 90\% coverage of the beam.
The neutron flux on the monochromator is calculated to be $5.4 \times 10^{9}$~cm$^{-2}$s$^{-1}$nm$^{-1}$ 
at 0.89~nm (at a reactor power of 58.3~MW). This value is based on a transmission simulation of the neutron guide, a gold foil activation measurement in front 
of the monochromator and spectral measurements made at the end of the neutron guide H171 \cite{Andersen/2010}.\footnote{The initial neutron guide H17 is splitt vertically into the guides H171 (reflectometer beam line \emph{Figaro}) and H172 (beam line for UCN production).}
Hence, assuming an integral neutron reflectivity of 40\% and the coverage of the monochromator one expects a
neutron flux of $2.0 \times 10^{9}$~cm$^{-2}$s$^{-1}$nm$^{-1}$ ($1.8 \times 10^{9}$~cm$^{-2}$s$^{-1}$nm$^{-1}$) at 0.89~nm and a reactor power of 58.3~MW (53.2~MW)
entering the converter vessel \cite{Wellenburg/2009a}.\footnote{The integral reflectivity is lower than the stated value of 50\% given in Ref.\ \cite{Wellenburg/2009a}, since the peak reflectivity of the used crystals was found to be smaller (60-70\% instead of 80\%). }

\subsection{Neutron wavelength spectrum and beam image}

\begin{figure}[tbp]
\centering
\includegraphics[width=0.45\textwidth]{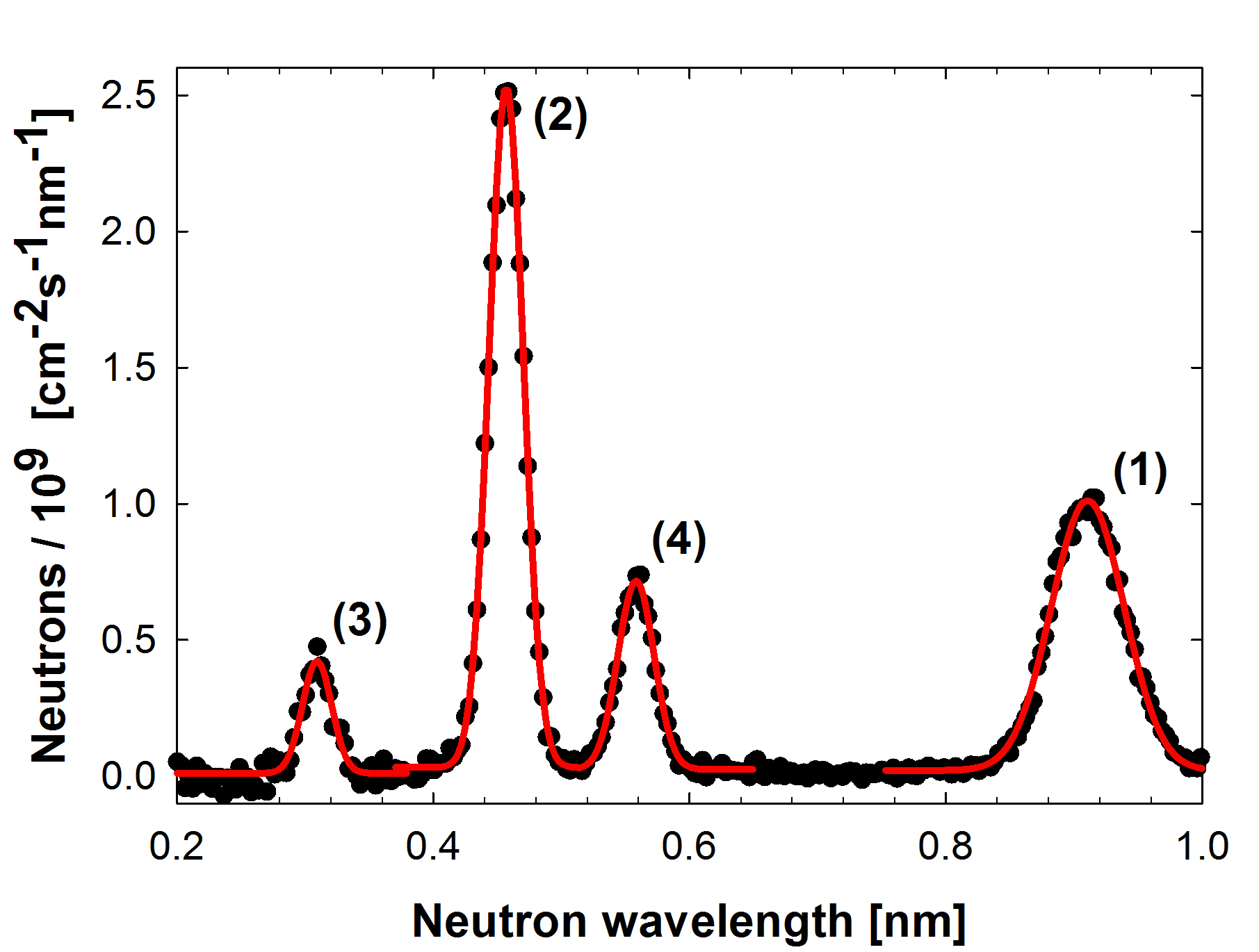}  
\caption{(Color online) Time-of-flight spectrum taken at the exit of the secondary neutron guide in the center of the beam 
(corrected for neutron scattering and absorption in air and normalized to a reactor power of 53.2 MW). 
The positions and widths of the four peaks have been fitted with independent Gauss functions (red lines): 
$\protect\lambda_1= (0.91\pm0.02)$~nm, $\sigma_1=0.0284(9)$~nm, $\protect\lambda_2=(0.46\pm0.01)$~nm,  
$\sigma_2=0.0134(4)$~nm, $\protect\lambda_3= (0.31\pm0.01)$~nm, $\sigma_3=0.0114(9)$~nm and 
$\protect\lambda_4=(0.56\pm0.01)$~nm, $\sigma_4=0.0142(5)$~nm. The given errors for the peak 
positions are dominated by the 2\% uncertainty in the wavelength calibration (compare text). The errors on
the peak width are a combination of statistical uncertainties and the wavelength calibration.}
\label{fig:TOF}
\end{figure}
\begin{figure}[tbp]
\centering
\includegraphics[width=0.45\textwidth]{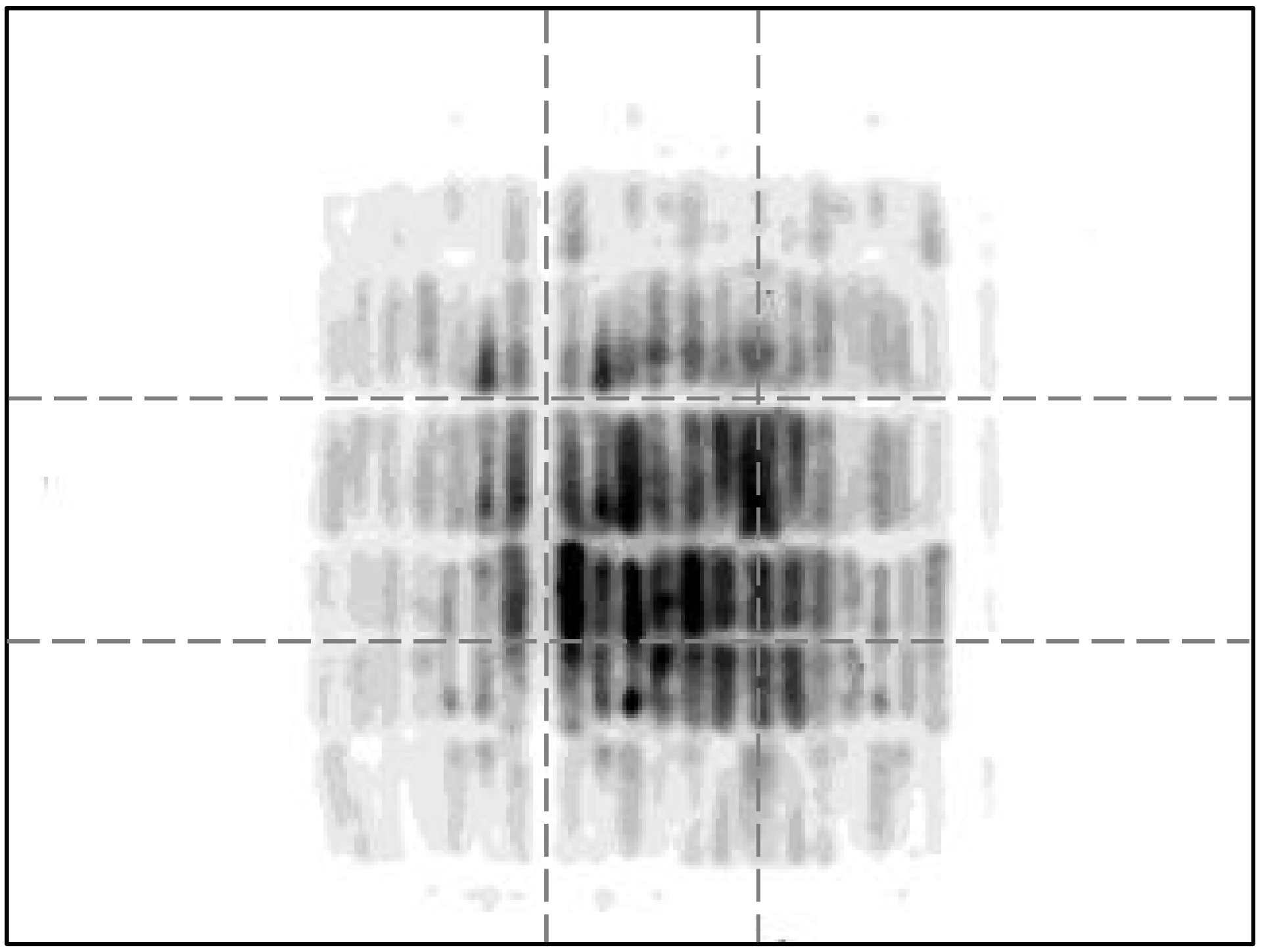}  
\caption{Pinhole image of the beam of neutron guide H172a ($320 \times 240$
pixel). Dark (light) pixel correspond to a high (low) neutron flux
intensity, i.e.\ a high (low) reflectivity of the crystals. The dashed lines
indicate the central square which represents the image of the direct beam. 
The eight surrounding images are due to mirror reflections in the secondary neutron guide.}
\label{fig:BeamImage}
\end{figure}

We evaluated the flux of the beam at the exit of the secondary guide 
by performing a gold foil activation analysis using a set of 13 gold foils equally 
distributed over the beam cross section. For the thermal equivalent flux we determined a mean 
value of $7.2 \times 10^8$~cm$^{-2}$s$^{-1}$ (at a reactor power of 53.2~MW) 
and the beam intensity was homogeneous over the whole cross section within the precision of the method.
In order to quantify the flux at 0.89~nm, the wavelength distribution
was measured in the center of the beam using a time-of-flight (TOF) setup. Its intrinsic
resolution was about $\Delta\lambda=0.01$~nm, due to the opening time of the
chopper and the detector thickness (1~cm). 
A correction due to neutron scattering and absorption in air was determined experimentally by 
comparison of two TOF spectra measured with the secondary guide (items 3, 4 and 5 in Fig.\ \ref{fig:beam}) 
vented with air and evacuated, respectively. The obtained wavelength dependent attenuation is 
about 15\% (11\%) per meter at 0.91~nm (0.46~nm). The corrected spectrum was normalized 
to the measured thermal equivalence flux and converted to the particle flux spectrum.
The result is presented in Fig.\ \ref{fig:TOF}. 
The plot shows four peaks: three of them ($\lambda_1$, $\lambda_2$ and $\lambda_3$) correspond to the
first three orders of Bragg reflections due to scattering on stage-2 type crystals and
one peak ($\lambda_4$) is caused by stage-1 contamination in the
crystals. The second order peak ($\lambda_2$) exceeds the first order
peak ($\lambda_1$) due to the shape of the cold neutron spectrum incident on
the monochromator exhibiting a broad maximum around 0.4~nm \cite{Andersen/2010}. \newline
This measurement was performed with a short chopper to detector distance to assure that the full
divergence of the different beam components is taken into account
(the TOF setup was replacing the short neutron guide section, item 5 in Fig.\ \ref{fig:beam}). 
In order to correct for position and timing offsets two TOF spectra were taken at 
distances of about 425~mm and 675~mm, each with an estimated systematic error of $2-3$~mm. 
This error results in an approximate 2\% uncertainty in the calibration of the 
wavelength axis. The determined peak position at $\protect\lambda_1= (0.91\pm0.02)$~nm, 
thus, agrees with the expected value of 0.89~nm within the given error. \newline
However, the spectrum also reveals a lack of intensity in the 
$\lambda_1$-peak with respect to the expected value presented previously.
Taking into account the broadening due to the intrinsic TOF resolution, one determines a 
flux of $(1.0 \pm 0.2) \times 10^{9}$~cm$^{-2}$s$^{-1}$nm$^{-1}$ at 0.89~nm and a reactor power of 53.2~MW. 
A 20\% uncertainty, quoted on this result, was assigned to account for possible small deviations of the 
apparent peak position from 0.89~nm and uncertainties in the gold foil activation measurement.
Normalized to equal reactor power, the differential flux is a factor 1.8 lower than expected.
As the reduced flux could be caused by a non-perfect alignment of the crystals, 
a pinhole image of the beam was taken using a standard neutron CCD-camera \cite{NeutronCamera/web}. 
The result is presented in Fig.\ \ref{fig:BeamImage}, showing that the individual crystals can
be easily identified due to the gaps between them. 
Observable is also a variation of the crystal quality (and/or alignment, which cannot be judged
from a single pinhole picture).
Furthermore, the entire crystal stack seems to be mounted several
millimeters too low, resulting in a white stripe with low intensity at the top of
the direct beam image. Hence, it appears likely that part of the missing intensity 
can be recovered by optimizing the alignment and reflectivity of the monochromator crystals, and by reducing the gaps between them.

\section{Production of ultracold neutrons}

\begin{figure}[tbp]
\centering
\includegraphics[width=0.45\textwidth]{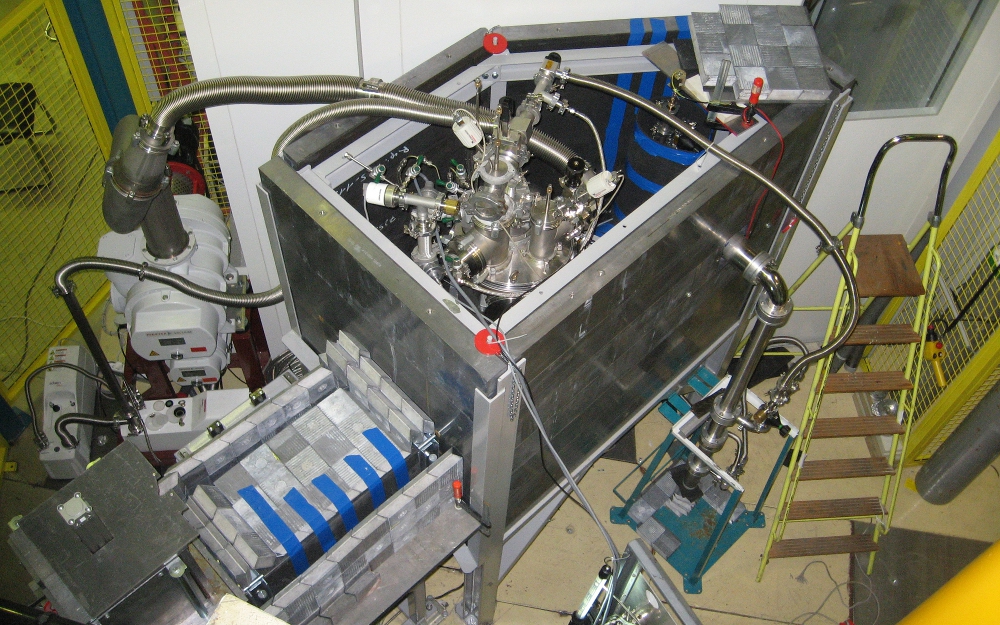}  
\caption{(Color online) Setup of the new UCN source at the beam line H172a. The cryostat
containing the superfluid helium converter is housed in a lead castle. Shown
is the setup with sidewards UCN extraction as used in the experiments
described here. The shutter for the cold neutron beam is situated in the
lower left corner of the picture.}
\label{fig:source}
\end{figure}
A photograph of the UCN source is presented in Fig.\ \ref{fig:source}. The
source was characterized by performing several measurements which are
described in this section. In these measurements the UCN are 
counted using a $^3$He gas detector (with an efficiency close to 100\%). It is
attached to the extraction guide made from stainless steel with a bend downward to
let the UCN gain gravitational energy to increase the transmission probability through
the aluminum entrance window of the detector (neutron optical
potential of 54~neV), as shown in Fig.\ \ref{fig:cryo} and visible on the
right in Fig.\ \ref{fig:source}. 

\subsection{UCN production as a function of temperature}

\begin{figure}
\centering
\subfigure[]{\includegraphics[width=0.45\textwidth]{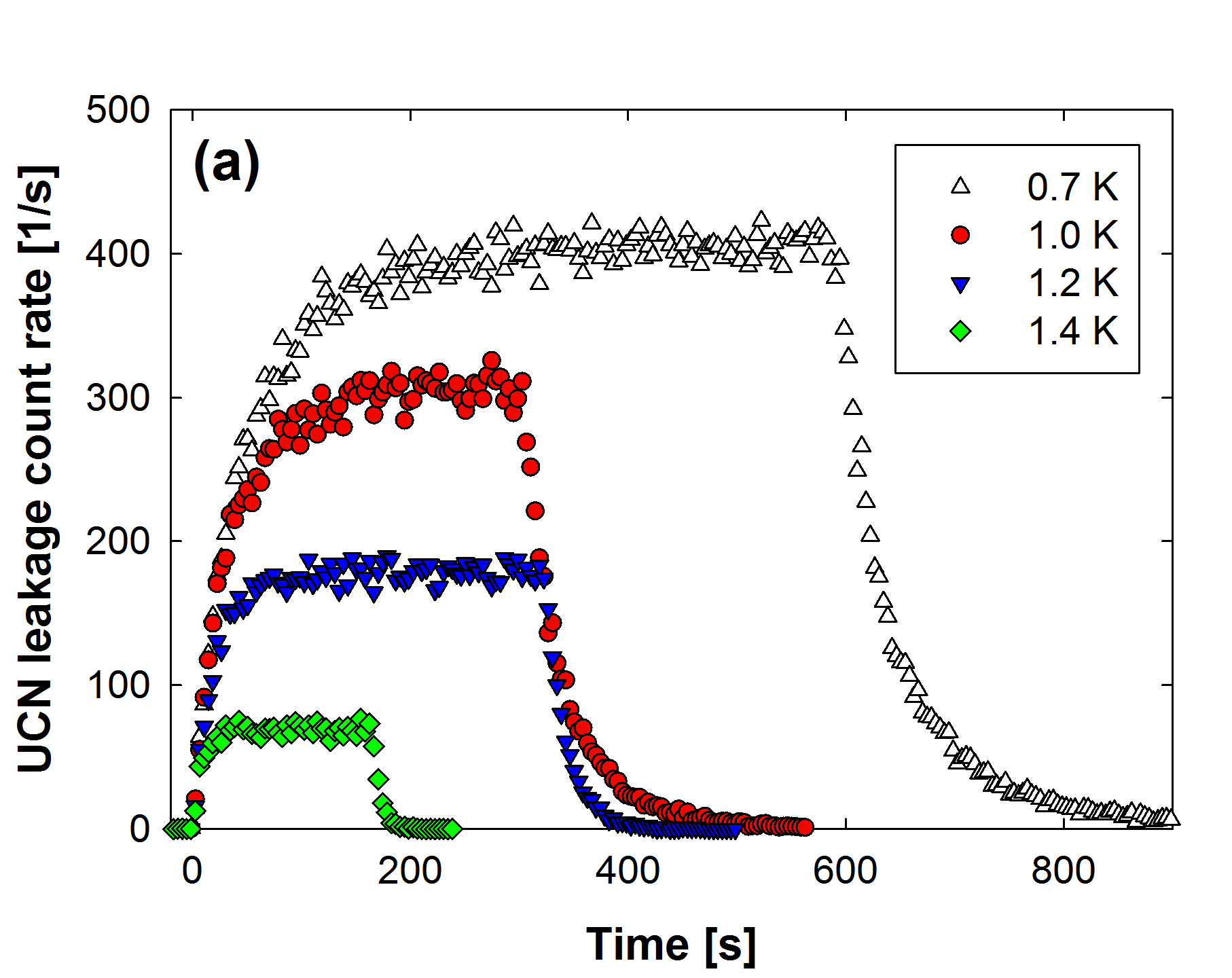}  } 
\subfigure[]{\includegraphics[width=0.45\textwidth]{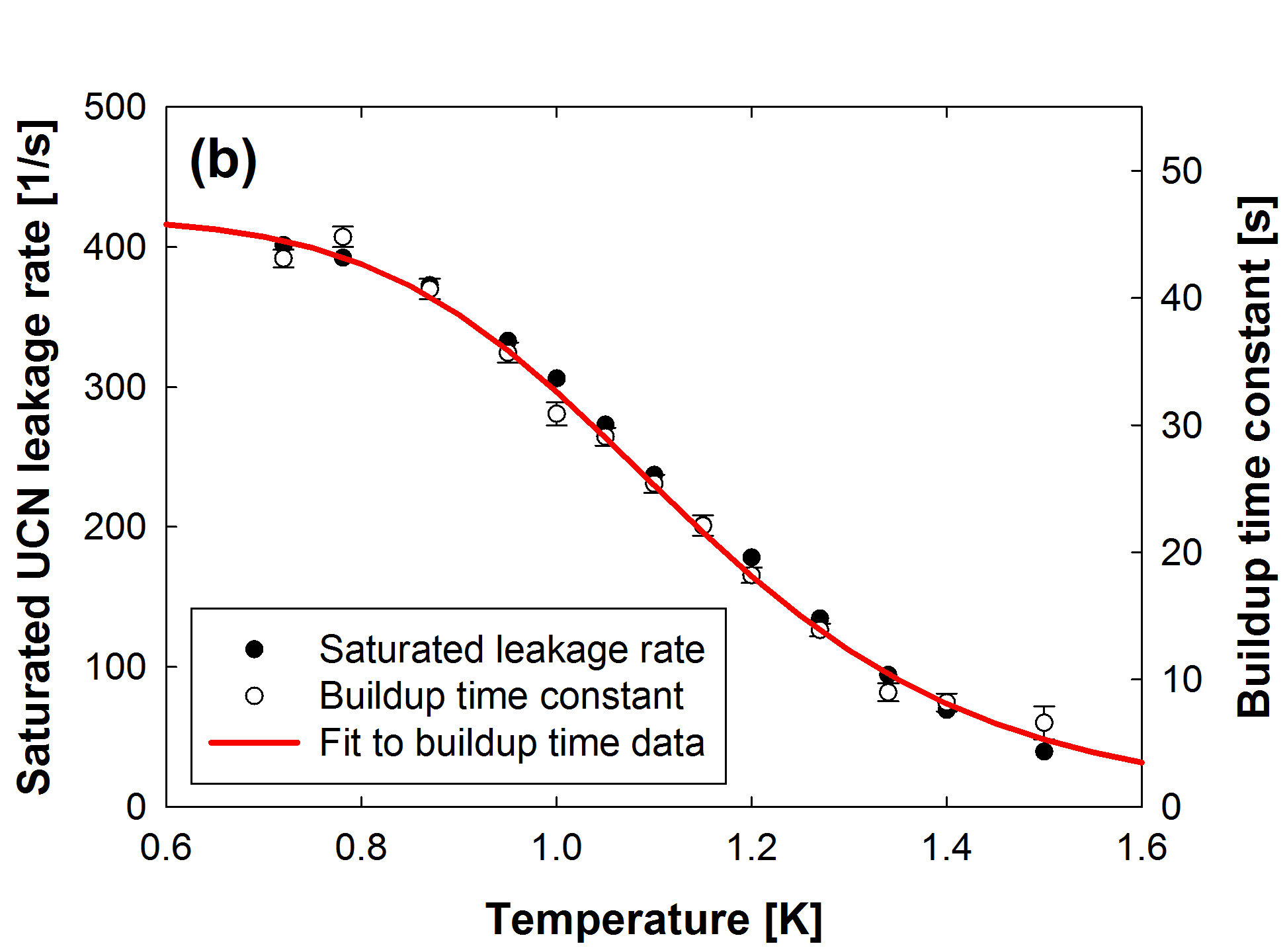}  }  
\caption{(Color online) (a) UCN leakage rate from the UCN storage volume to the detector
with a slightly opened UCN shutter valve for different temperatures of the
superfluid $^4$He. The converter vessel is irradiated with cold neutrons for
different durations with starting time set to zero.  For clarity 
4 neighboring data points have been averaged and recombined to one. (b) Saturated UCN
leakage rate through the slightly opened shutter valve and the corresponding
buildup time constant for different $^4$He temperatures. The latter are
obtained from a fit to the exponential rise of the data points in (a).
The line represents a fit to the data points of the buildup time (compare text).}
\label{fig:tempdepUCNproduction}
\end{figure}

The cooldown of the superfluid $^{4}$He from 1.5~K to the
base temperature of 0.7~K was slow enough to follow the increase of the UCN
density in the storage volume as a function of temperature. For that purpose
the UCN shutter valve was slightly opened providing a small slit for the UCN
to leak from the production volume to the detector. Fig.\ \ref{fig:tempdepUCNproduction}a 
depicts the buildup and decay of the UCN leakage
rate measured at four different temperatures as a function of time after
opening and closing the cold neutron shutter, i.e.\ irradiation of the
converter vessel with cold neutrons. The rise of the count rates to
a plateau is well described by a single exponential buildup. 
The fit parameters "saturated UCN leakage rate" and "buildup time constant" $\tau_{\text{b}}$ 
are presented as a function of temperature $T$ in Fig.\ \ref{fig:tempdepUCNproduction}b. 
The plot demonstrates the expected proportionality between these two parameters.
Compared to a separate measurement, where the buildup time constant with a completely closed shutter valve was determined
to be $(67\pm 3)$~s at 0.7~K (compare below), here, the buildup time reaches a saturation value of 46~s. 
This indicates that the storage lifetime of the UCN in the storage volume is limited by the slightly opened shutter valve. \\
The temperature dependence of the buildup time constant was fitted using the function: 
\begin{equation}
  \frac{1}{\tau_{\text{b}}(T)} = \frac{1}{\tau_0} + \frac{1}{\tau'} \cdot T^c  
\end{equation}
where $T$ is the $^4$He temperature in units of Kelvin and with the fit parameters $\tau_0$, $\tau'$ (time constants) and $c$.
A least squares fit to the data points yields: $\tau_0 = (46 \pm 1)$~s, $\tau'=(111 \pm 7)$~s and $c=7.2 \pm 0.3$.
Here, $\tau_0$ is specific for our UCN storage volume with the slightly opened UCN shutter valve, while the storage time constant in $^4$He  at 1~K, $\tau'$, and the exponent $c$ generally
describe the temperature dependent up-scattering of UCN in liquid $^4$He. The obtained values are in agreement 
with the findings given in Ref.\ \cite{Golub/1983, Yoshiki/1992} and with the two-phonon up-scattering term derived in Ref.\ \cite{Golub/1979}. 
The corresponding fit curve is presented in Fig.\ \ref{fig:tempdepUCNproduction}b. \newline
In a measurement with fully opened UCN valve
performed at 1.27~K we obtained a continuous UCN rate of 1400~s$^{-1}$
through the extraction hole with area 4~cm$^{2}$.

\subsection{Storage lifetime measurements and UCN density}
\label{sec:delextract}

\begin{figure}
	\centering
		\includegraphics[width=0.45\textwidth]{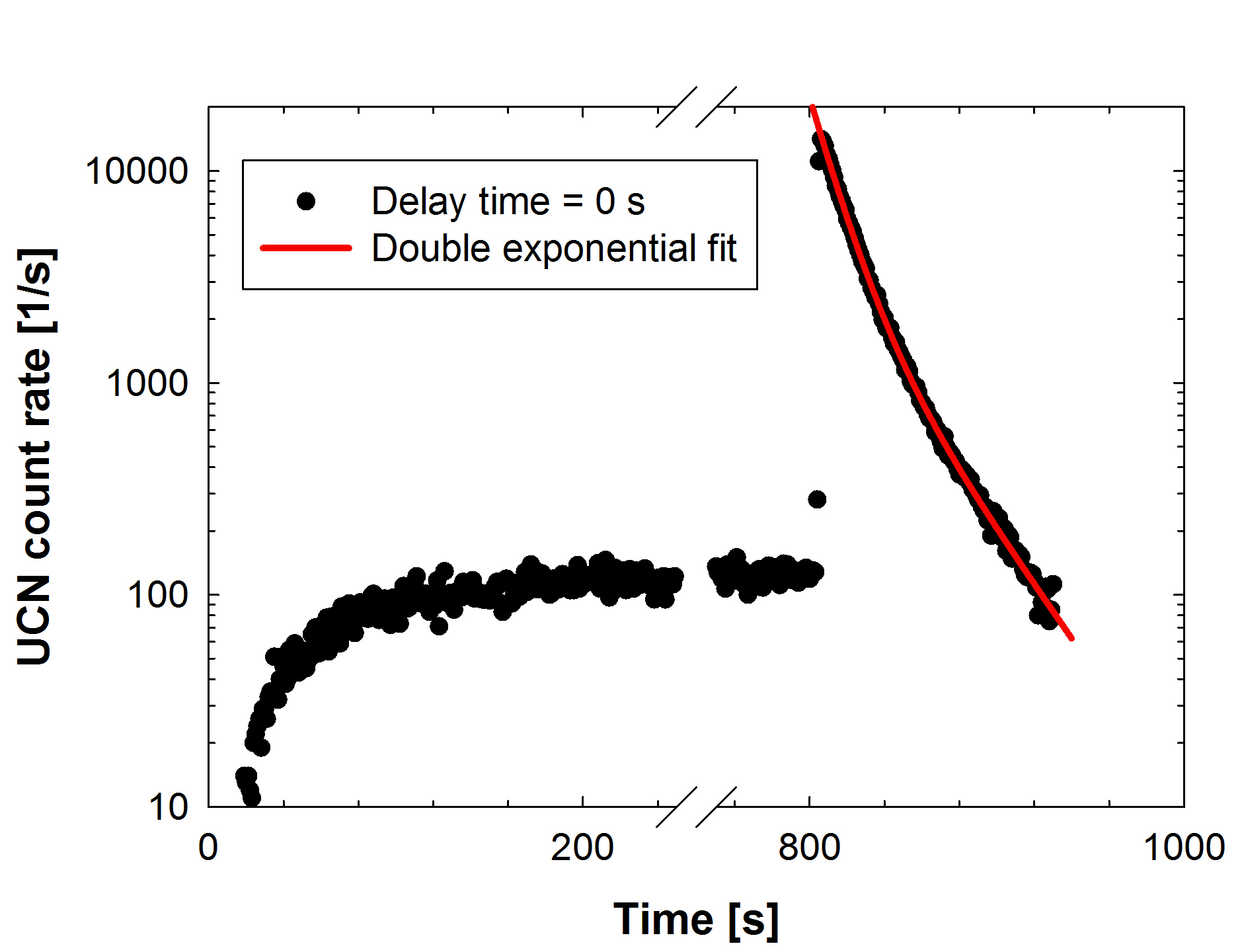}
	\caption{(Color online) Time plot of a delayed extraction measurement with zero delay time. 
	Here, the UCN shutter valve is opened after approximately 800~s of irradiation 
	with cold neutrons and accumulation of UCN. The total number of detected neutrons 
	during the 125~s long extraction is approximately $2.74 \times 10^5$. The double exponential 
	fit to the extraction data yields two time constants: $(13 \pm 1)$~s and $(34 \pm 1)$~s.}
	\label{fig:superspill}
\end{figure}

The storage lifetime of the UCN in the production volume is limited by several loss mechanisms: 
(\emph{i})    the neutron $\beta$-decay with $\tau_{\beta} \approx 880$~s,
(\emph{ii})   phonon up-scattering processes in liquid $^4$He,
(\emph{iii})  absorption by a residual $^3$He contamination in the liquid helium,
(\emph{iv})   absorption and up-scattering due to wall collisions of the neutrons,
(\emph{v})    absorption and up-scattering by impurities, and
(\emph{vi})   losses through gaps in the storage volume wall. \\
As determined above the effect due to the up-scattering in $^4$He can be approximated at low temperatures by 
$\tau_{\text{up}} \approx 110~\text{s} / T^7$, where $T$ is in units of Kelvin. 
For 0.7~K this yields about 1350~s. 
Neutron losses due to wall collisions depend on the neutron velocity and 
the mean free path given by $\bar{l}=4V/A$, where $A$ and $V$ are the total surface 
area and volume of the storage volume (here: $A \approx 0.3$~m$^2$, $V \approx 5$~liter and $\bar{l} \approx 7$~cm). 
For estimates, we employ the maximum neutron velocity $v_{\text{max}}$ and the maximum loss 
probability per wall collision $\mu_{\text{max}} \equiv \pi f$, with $f$ being the so-called UCN loss factor 
(for BeO: $\mu_{\text{max}} \approx 4 \times 10^{-5}$) \cite{Golub/1991}.
From the kinetic UCN gas theory it follows that $\tau_{\text{wall}}=\bar{l} / (v_{\text{max}} \cdot \mu_{\text{max}})$.
Hence, for the maximal UCN velocity of 6~m/s, 
limited by the neutron optical potential of the vertical stainless steel guide, 
this yields $\tau_{\text{wall}} \approx 300$~s. 
Taking only these two loss mechanisms and the $\beta$-decay into account,
one obtains an estimated upper limit for 
the storage lifetime in our UCN production volume of $\tau^* \approx 200$~s.  \\
In order to determine the storage lifetime and time constants 
for the UCN extraction, we performed delayed UCN extraction experiments.
In these measurements the converter vessel is irradiated with cold neutrons
for a period much longer than the buildup time constant, while the
UCN shutter valve remains closed. The UCN valve is then opened with a
delay time after closing the beam. 
\begin{figure}[tbp]
\centering
\subfigure[]{\includegraphics[width=0.45\textwidth]{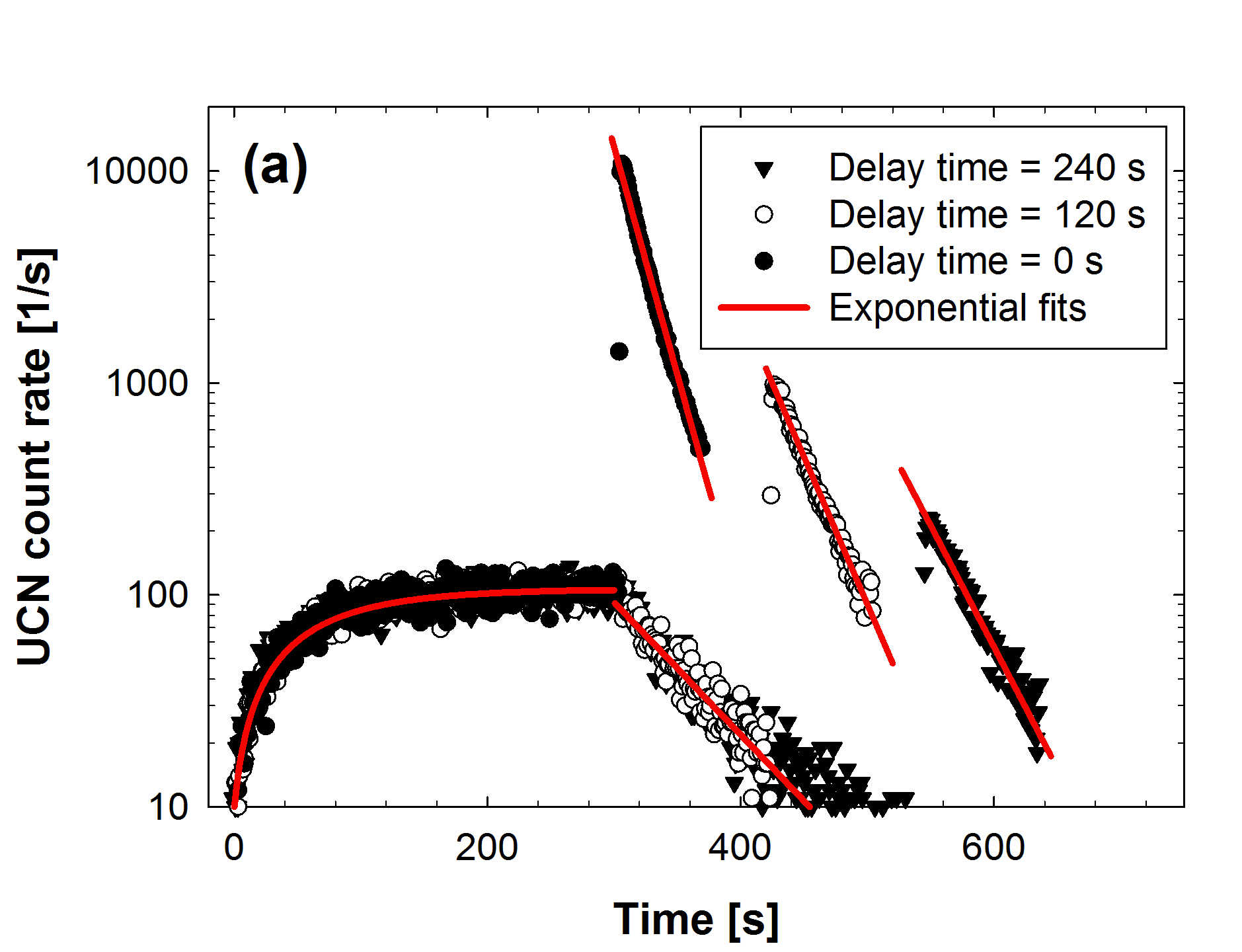}  } 
\subfigure[]{\includegraphics[width=0.45\textwidth]{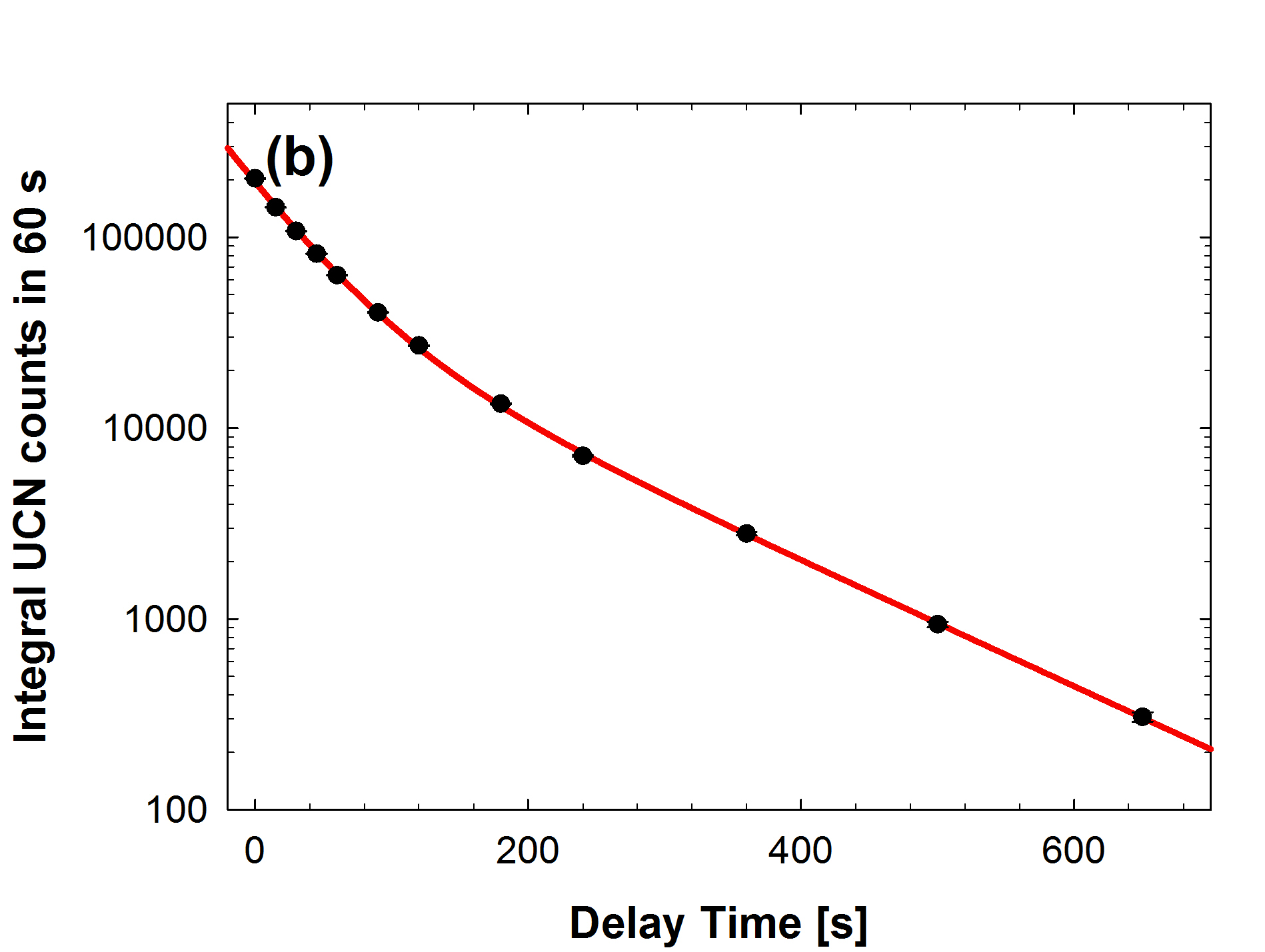}  } 
\subfigure[]{\includegraphics[width=0.45\textwidth]{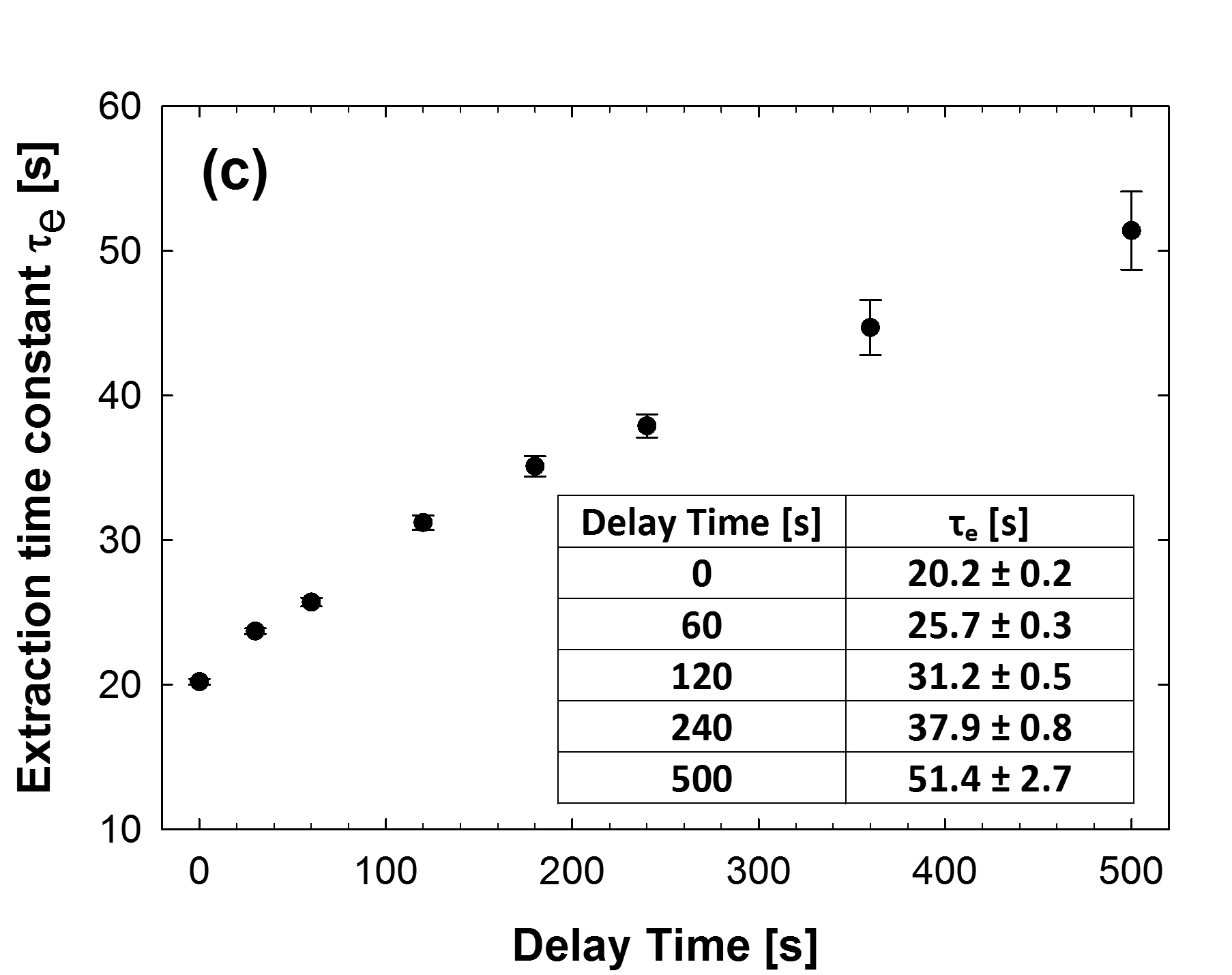}  }  
\caption{(Color online) (a) Time plots of the delayed extraction measurements at a $^4$%
He temperature of about 0.7~K. The plot depicts the detected UCN count rate
as a function of time and corresponding single exponential fit curves (red
lines). The converter vessel is irradiated with the cold neutron beam for a
300~s long UCN accumulation period. After the irradiation and a variable
delay time the UCN shutter valve is opened and the UCN are extracted towards
the detector. (b) The integral counts of the extracted UCN as a function of
the delay time. The line shows the fit with a double exponential decay function. 
(c) Extraction time constant $\tau_{\text{e}}$ as a function of the delay time, obtained from a
single exponential decay fit to the data points in (a) during emptying of
the storage volume.}
\label{fig:delayedextraction}
\end{figure}
In Fig.\ \ref{fig:delayedextraction}a typical time plots of the UCN count rate
are shown for the superfluid $^4$He held at 0.7~K. During the 300~s long
irradiation with cold neutrons the buildup of UCN density can be followed
due to the leaking of UCN through the fully closed shutter valve. After
closing the beam the density decays with a time constant in agreement with
the buildup time constant. An exponential fit to the data points yields 
$(67 \pm 3)$~s for the buildup and $(70 \pm 2)$~s for the decay time constant.
After the delayed opening of the shutter valve the accumulated UCN are
emptied from the source to the detector. The integral UCN counts in such an emptying curve
are presented in Fig.\ \ref{fig:delayedextraction}b as a function of
the delay time. Here, the integral counts are summed over the first 60~s
after opening the shutter valve. A double exponential fit to the data points
yields two time constants: $\tau_{\text{1}}=(43 \pm 1)$~s and $\tau_{\text{2}%
}=(132 \pm 2)$~s, with approximately 78\% of the measured intensity in the
faster component. The presence of rather different time constants is
characteristic for a broad UCN energy spectrum stored in a material bottle,
since the frequency of wall collisions and the average loss probability per
wall collision are strongly energy dependent. The resulting effect of
spectral shaping of the trapped UCN is well visible in the impressive rise
of extraction time constants with increasing delay time presented in Fig.\ \ref{fig:delayedextraction}c.\footnote{Note that for long delay times, due to the longer extraction time constants, the summation of neutron counts over the first 60~s after opening the UCN valve yields an underestimation of the neutron count rate and consequentely of $\tau_2$.} \\
On the other hand, the UCN absorption by residual $^3$He in the liquid helium does not depend on the
neutron velocity and, thus, we can provide an upper limit for the isotopical purity of $^3$He/$^4\text{He} <  10^{-10}$.
This limit is obtained by attributing the full difference between the estimated storage lifetime 
$\tau^*$ and the measured longer time constant $\tau_{\text{2}}$ to the $^3$He contamination 
via $1/\tau_{\text{$^3$He}} = 1 / \tau_{\text{2}} - 1/ \tau^* \approx (400\text{~s})^{-1}$,
and employing the thermal neutron absorption cross section of $^3$He of 5330~barn.\footnote{The $^3$He number density can be estimated by $\rho_{\text{$^3$He}}=(\tau_{\text{$^3$He}} \cdot 5330\text{~barn} \cdot 2200\text{~m/s})^{-1}$.} \\
In conclusion, the experimentally determined buildup and decay time constants are 
about a factor 3 smaller than $\tau^*$. 
This can be attributed to losses through gaps, for instance between the individual 
BeO ceramic pieces (e.g.\ a relatively large combined slit area of 0.3~cm$^2$ 
would lead to an additional loss channel with a time constant of about 100~s), 
and losses due to absorption or up-scattering by possible impurity depositions on the walls of the production volume.\\
An extraction measurement with a longer emptying period is presented in Fig.\ \ref{fig:superspill}.
From this a lower limit for the UCN density in the production volume can be determined, 
not taking losses due to an imperfect UCN guide system into account.
In this measurement a total number of $2.74 \times 10^5$ UCN was counted, corresponding to a density of at least 55~cm$^{-3}$. 

\subsection{Duty cycle of the UCN source}

\begin{figure}[tbp]
\centering
\includegraphics[width=0.45\textwidth]{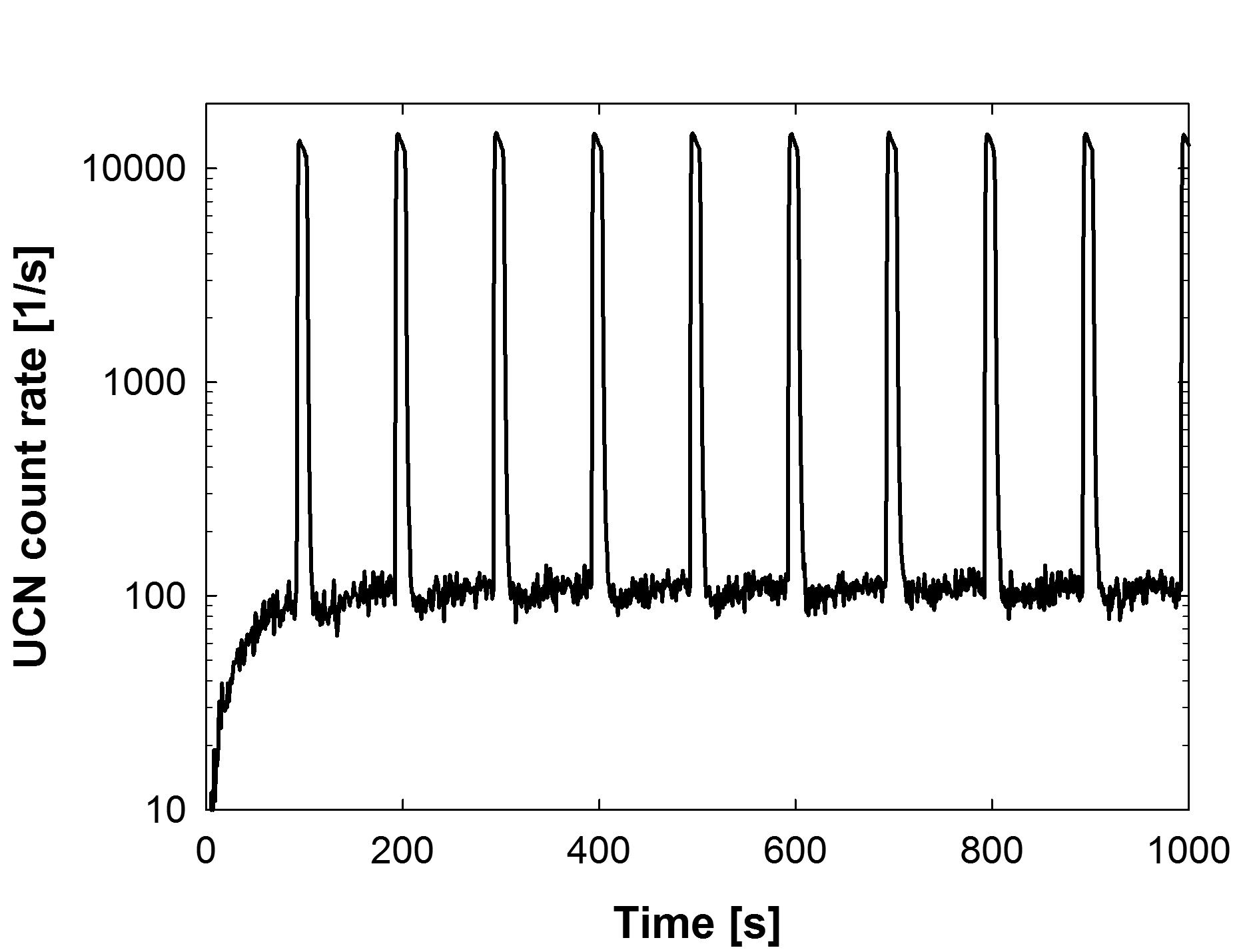}  
\caption{Repeated extraction of UCN during permanent irradiation of cold
neutrons with a duty cycle of 10\% (defined as fraction of time with UCN
shutter valve open per accumulation-extraction cycle). In each emptying curve a total
amount of approximately $1.4 \times 10^5$ UCN is extracted (shutter valve open for 10~s, counting during 20~s after start of shutter opening).}
\label{fig:dutycycle010}
\end{figure}

In the presented experiments, we observed no significant heating of the
source due to the irradiation with the neutron beam. On the other hand,
after opening the UCN shutter valve the temperature of the superfluid $^{4}$%
He increases due to thermal radiation entering the bath along the extraction
guide. 
However, as most UCN experiments involve filling experimental bottles in 
a periodic manner (e.g.\ neutron lifetime and electric dipole moment experiments), 
the valve needs to be open only for short moments between much longer 
periods of manipulation or storage of the extracted UCN. The UCN 
density in the source can be replenished during these periods.
To study a typical operation of the source we repeated UCN 
extractions with the neutron beam switched on all the time. 
In Fig.\ \ref{fig:dutycycle010} measured UCN counts for a
duty cycle of 10\% are shown, where in each cycle the shutter valve is kept open for
10~s, followed by a UCN accumulation time of 90~s. The total number of
counts in each emptying curve was approximately $1.4 \times 10^5$. Within 1~h of operation the $%
^{4}$He temperature rose from 0.7~K to 0.9~K, associated with a small drop
in UCN count rate. Hence, without provision of additional cooling power, a
slightly reduced duty cycle should be applied in order to maintain a
constant UCN output. For a duty cycle of 5\% the mean temperature of the helium
bath stayed at 0.7~K without any observable drift \cite{Zimmer/2011}.  

\subsection{Comparison with expected UCN density}
\label{sec:compareUCNdensity}


For an estimate of the UCN density $\rho _{\text{UCN}}$ to be expected for this source, we 
neglect energy dependences and write $\rho _{\text{UCN}}=p\tau $ with an ensemble average value for $\tau $,
for which we use $(67\pm 3)$~s as determined from the UCN buildup
measurements through the leaking UCN valve. The production rate density $%
p=\int_{0}^{U}p\left( \epsilon \right) \mathrm{d}\epsilon $ is defined as
the conversion rate of cold neutrons to UCN with kinetic energies smaller
than the trap potential depth $U$ (accounting for the neutron optical potential of superfluid $^4$He of 19~neV). 
For our converter vessel, the wall material with
the smallest neutron optical potential is uncoated stainless steel
of the vertical UCN guide section and the shutter valve (items 17 and
18 in Fig.\ \ref{fig:cryo}). Hence, from $p\propto U^{3/2}$ and using the
value calculated for a vessel made of beryllium \cite{Wellenburg/2009} one derives $%
p=(3.0\pm 0.2)\times 10^{-9}\ \text{d}\phi /\text{d}\lambda \mid _{\lambda
^{\ast }}$~cm$^{-3}$s$^{-1}$, with the differential neutron flux at $\lambda
^{\ast }=0.89$~nm given in cm$^{-2}$s$^{-1}$nm$^{-1}$. Since the vessel's
side walls from sintered BeO ceramics do not guide the cold neutron beam,
this value needs to be divided by a factor of $1.4 \pm 0.1$ to obtain
the spatially averaged production rate density, where the stated value is
based on simple geometrical considerations. For the neutron flux as
obtained from the TOF measurements, we thus expect $\rho _{\text{UCN}}\approx (140\pm 20)$~cm$^{-3}$. 
On the other hand, the experimental value of 55~cm$^{-3}$ quoted
before was exclusively based on detected UCN. For a valid comparison one
needs to apply a correction by the ratio $\left( \tau + \tau_{\text{e}}\right)/ \tau
\approx $ $1.3$ to account for UCN losses within the converter
during extraction with an extraction time constant of $\tau_{\text{e}}\approx 20$~s,
i.e.\ $\rho _{\text{UCN}}^{\text{exp}}\approx 70$~cm$^{-3}$. The agreement looks
not unreasonable since we did not take into account the transmission of 
the UCN extraction guide with its two 90° bends, the transmission of the 
detector entrance window and apertures for pumping.

\section{Conclusion and Outlook}
\label{sec:conclusion}

The installation of the apparatus described here represents a milestone towards a "next
generation" UCN source. The measured UCN density exceeds that of the long
standing user source PF2 at ILL. However, there is still large room for
improvements, and it is instructive to consider identified factors $f_{i}$ 
that would result in a significant gain in UCN density if the following measures are taken.
A first gain will result from a coating 
of all parts inside the production volume and of the UCN guides with beryllium or any other material with
high neutron optical potential and low absorption, such as diamond-like
carbon \cite{Grinten/1999,Atchison/2006,Atchison/2007} or cubic $^{11}$BN 
\cite{Sobolev/2010}. 
For beryllium the increased trap depth with respect to the vessel 
partly made of stainless steel, will yield a gain factor $f_{1}=1.67$.
Second, for the given geometry, an effective storage time constant of approximately 200~s 
should be feasible (compare $\tau^*$).\footnote{Note that the Cryo-EDM collaboration, 
with a superthermal source consisting of a $3$~m long tubular vessel with $63$~mm diameter made of 
beryllium-coated copper and closed off by beryllium windows, but without extraction of UCN through a
liquid helium-vacuum interface, has measured a storage time of 160~s \cite{Grinten/2009}.}
The improved storage lifetime would correspond to a gain factor $f_{2}=3$. 
Third, the mentioned loss due to beam divergence can be avoided if
the side walls of the converter vessel guide the cold neutron beam, i.e.\ 
$f_{3}=1.4 \pm 0.1$. This can be achieved by sputtering a supermirror coating 
on the production volume walls before depositing the UCN reflecting top layer.
Neglecting any further gain due to improved UCN extraction
efficiency, the saturated UCN density in the source installed
at its present beam location H172a with characteristics as described in this
paper would thus become $f_{1}f_{2}f_{3}\cdot 70$~cm$^{-3}=\left( 490\pm
40\right)$~cm$^{-3}$. \\
In the meantime a new secondary beam position, H172b, with similar characteristics has been put in operation. This beam line will be used to bring the technique of superthermal UCN
production in superfluid $^{4}$He to maturity using a successor of the UCN source cryostat presented here. 
Since the  monochromation losses were found to be significantly larger than initially anticipated, another large gain could be achieved
by omission of the monochromator and implementation of the source in the direct/white beam at the exit of the guide H172, where the differential neutron flux at 0.89~nm is
about a factor 5 higher than in the secondary beam \cite{Andersen/2010}.

\section{Acknowledgment}
\label{sec:ack}

The authors gratefully acknowledge help and support by D. Berruyer, B. van den Brandt, T. Brenner, M. Guigou, S. Mironov, and the entire GRANIT collaboration.

\appendix

\end{document}